\documentclass[rmp,twocolumn,showkeys]{revtex4-1}
\usepackage{amsmath}
\usepackage{amssymb}
\usepackage{graphicx}
\usepackage[pdftex]{hyperref}
\usepackage{float}
\newcommand{\He}{$^{3}$He}
\newcommand{\Hea}{$^{3}$He-A}
\newcommand{\Heb}{$^{3}$He-B}

\newcommand{\nm}{\,\mbox{nm}}
\newcommand{\mbar}{\,\mbox{bar}}
\newcommand{\angstrom}{\textup{\AA}}

\newcommand{\pPCP}{p_{\mbox{\tiny PCP}}}
\newcommand{\betawc}[1]{\beta_{#1}^{\text{wc}}}
\newcommand{\betasc}[1]{\beta_{#1}^{\text{sc}}}
\def\nicefrac#1#2{\genfrac{}{}{}{1}{#1}{#2}}
\def\text#1{\mbox{\tiny #1}}
\def\ns{\negthickspace}

\newcommand{\wpreB}{\frac{\gamma^{2}}{\chi_{B}}g_{D}}

\newcommand{\Dzz}{\left\langle A_{zz}^2 \right\rangle}
\newcommand{\Dxx}{\left\langle A_{xx}^2 \right\rangle}
\newcommand{\Dyy}{\left\langle A_{yy}^2 \right\rangle}
\newcommand{\Dxz}{\left\langle A_{xz}^2 \right\rangle}
\newcommand{\Dzx}{\left\langle A_{zx}^2 \right\rangle}
\newcommand{\DxxMDyy}{\left\langle (A_{yy}-A_{xx})^2 \right\rangle}
\newcommand{\DxxPDyy}{\left\langle (A_{yy}+A_{xx})^2 \right\rangle}

\newcommand{\DxxDzz}{\left\langle A_{xx}A_{zz} \right\rangle}

\newcommand{\vz}{{\bf z}}
\newcommand{\vH}{{\bf H}}
\setcitestyle{numbers,square}
\begin{document}
\title{Strong-coupling and the Stripe phase of $^3$He}
\author{Joshua J. Wiman$\,$}
\email{jjwiman@u.northwestern.edu}
\author{J. A. Sauls$\,$}
\email{sauls@northwestern.edu}
\affiliation{Department of Physics \& Astronomy, Northwestern University}
\date{\today}
\begin{abstract}
Thin films of superfluid \He{} were predicted, based on weak-coupling BCS theory, to have a stable phase 
which spontaneously breaks translational symmetry in the plane of the film.
This crystalline superfluid, or ``stripe'' phase, develops as a one dimensional periodic array of domain 
walls separating degenerate B phase domains.
We report calculations of the phases and phase diagram for superfluid \He{} in thin films using a 
strong-coupling Ginzburg-Landau theory that accurately reproduces the bulk \He{} superfluid phase diagram.
We find that the stability of the Stripe phase is diminished relative to the A phase, but the Stripe phase
is stable in a large range of temperatures, pressures, confinement, and surface conditions.
\end{abstract}
\maketitle
\section{Introduction}\label{intro}
The theoretical prediction of a crystalline superfluid, or ``stripe''  phase, that spontaneously 
breaks translational symmetry in thin films of \He{} \cite{vor07}, along with advances in nanoscale fabrication 
and experimental instrumentation \cite{lev13}, has renewed interest in the properties of superfluid \He{} in thin 
films and confined geometries.
In the weak-coupling limit of BCS theory the Stripe phase is predicted to be stable in a 
large region of temperature and pressure for films of thickness $D\sim 700\nm{}$.
However, recent experiments on \He\ confined in slabs of thickness $D\approx 700\nm{}$ 
and $D\approx 1080\nm{}$ have failed to detect evidence of the Stripe phase \cite{lev13}.

A limitation of the Vorontsov and Sauls theory is that it does not include strong-coupling corrections
to the BCS free energy.
In bulk \He{}, weak-coupling theory predicts a stable B phase at all temperatures and pressures; 
however, the A phase is found to be stable experimentally at $T_c$ and pressures above 
$\pPCP{} \approx 21.22 \mbar{}$, with a first-order transition at $T_{\text{AB}}<T_c$ to the B phase.
Theoretically accounting for the stability of the A phase requires including next-to-leading order 
corrections to the full free energy functional, i.e. corrections to the weak-coupling 
functional \cite{rai76}.
While these strong-coupling corrections are largest at high pressures, they remain significant 
even for $p \sim 0 \mbar{}$\cite{cho07}. Thus, for superfluid \He{} confined within 
a film, it is to be expected that strong-coupling effects will increase the stability of the 
A phase relative to both the B- and Stripe phases, which could diminish, or even eliminate, the 
experimentally accessible region of the Stripe phase.

In this paper we report our study of the A-Stripe and Stripe-B superfluid transitions using a
Ginzburg-Landau (GL) functional that incorporates strong-coupling corrections to the weak-coupling GL 
material coefficients and accurately reproduces the bulk superfluid \He{} phase diagram \cite{wim15}.
Within this strong-coupling GL theory we calculate the superfluid order parameter and phase diagram 
as a function of pressure, temperature, confinement, and surface conditions.

\section{Ginzburg-Landau Theory}\label{sec:1}

The general form of the p-wave, spin triplet order parameter for \He\ is given by the mean-field 
pairing self energy, which can be expanded in the basis of symmetric Pauli matrices ($S=1$) and 
vector basis of orbital momenta ($L=1$), 
\begin{equation}
\hat{\Delta}(\hat{p}) = \sum_{\alpha i} A_{\alpha i}\, (i \sigma_\alpha \sigma_y)\, \hat{p}_i \,,
\end{equation}
where $\hat{p}$ is the direction of relative momentum of the Cooper pairs defined on the Fermi surface,
and $A_{\alpha i}$ are the elements of a $3\times 3$ complex matrix,
\begin{equation}
A = 
\begin{pmatrix} 
A_{xx} & A_{xy} & A_{xz} \\
A_{yx} & A_{yy} & A_{yz} \\
A_{zx} & A_{zy} & A_{zz} 
\end{pmatrix}
\,,
\end{equation}
that transforms as a vector under spin rotations (with respect to $\alpha$) and (separately) as a 
vector under orbital rotations (with respect to $i$). We choose aligned spin and orbital coordinate axes.

\vspace*{8mm}
\subsection{Free energy functional}\label{subsec:1:1}

To determine the order parameter and the phase diagram of \He\ in a film geometry, we 
solve the Euler-Lagrange equations of the Ginzburg-Landau functional, subject to relevant boundary
conditions, and calculate the order parameter and the stationary free energy.
The GL functional is defined by bulk and gradient energies with temperature dependent strong-coupling 
corrections, and is supplemented by boundary conditions that we can tune from maximal to minimal 
pair-breaking \cite{wim15}.

The Ginzburg-Landau free energy functional is expressed in terms of invariants constructed from the 
order parameter matrix, $A$, and is given by \cite{thu87}
\begin{widetext}
\begin{align}
\Omega[A] 
&=
\int_{V}d\vec{R}\;
\left\{\vphantom{\frac{1}{3}} 
\alpha(T) Tr\left(A A^{\dagger}\right) 
+\beta_{1} \left|Tr(A A^{T})\right|^{2}
+\beta_{2} \left[Tr(A A^{\dagger})\right]^{2}  
\right. 
\\
& \qquad \left.
\vphantom{\frac{1}{3}} 
+\beta_{3}\, Tr\left[A A^{T} (A A^{T})^{*}\right] 
+\beta_{4}\, Tr\left[(A A^{\dagger})^{2}\right]
+\beta_{5}\, Tr\left[A A^{\dagger} (A A^{\dagger})^{*}\right] 
\right. 
\nonumber\\
& \qquad \left.
\vphantom{\frac{1}{3}} 
+K_{1} \left(\nabla_{k}A_{\alpha j} \nabla_{k}A_{\alpha j}^{*}\right)
+K_{2} \left(\nabla_{j}A_{\alpha j} \nabla_{k}A_{\alpha k}^{*}\right)
+K_{3} \left(\nabla_{k}A_{\alpha j} \nabla_{j}A_{\alpha k}^{*}\right) 
\right\}
\,.
\nonumber
\end{align}
\end{widetext}

In the weak-coupling limit the GL material parameters are given by
\begin{eqnarray}
\alpha^{\text{wc}}(T) 
&=& 
\nicefrac{1}{3}N(0)(T/T_{c}-1)\,,
\\
2\beta_{1}^{\text{wc}} 
&=& -\beta_{2}^{\text{wc}} = -\beta_{3}^{\text{wc}} = -\beta_{4}^{\text{wc}} = \beta_{5}^{\text{wc}}
\,,
\label{eq-beta-wc}
\\
\beta_{1}^{\text{wc}}
&=&
-\frac{N(0)}{(\pi k_{\text{B}}T_{c})^2}\left\{\frac{1}{30}\left[\frac{7}{8}\zeta(3)\right]\right\}
\,,
\\
K_{1}^{\text{wc}} 
&=&
K_{2}^{\text{wc}} = K_{3}^{\text{wc}} = \frac{7 \zeta(3)}{60} N(0)\, \xi_{0}^2
\,,
\end{eqnarray}
and determined by the normal-state, single-spin density of states at the Fermi energy, $N(0)$, the bulk 
transition temperature, $T_c$, and the Fermi velocity, $v_f$.
The Cooper pair correlation length $\xi_{0} \equiv \hbar v_{f} / 2\pi k_{B} T_{c}$ varies from 
$\xi_{0}\simeq 770\,\angstrom$ at $p=0\,\mbox{bar}$ to $\xi_{0}\simeq 160\,\angstrom$ at $p = 34\,\mbox{bar}$.

\subsection{Strong-coupling corrections}\label{subsec:1:2}

\begin{figure}[t]
\begin{center}
\includegraphics[width=1\linewidth]{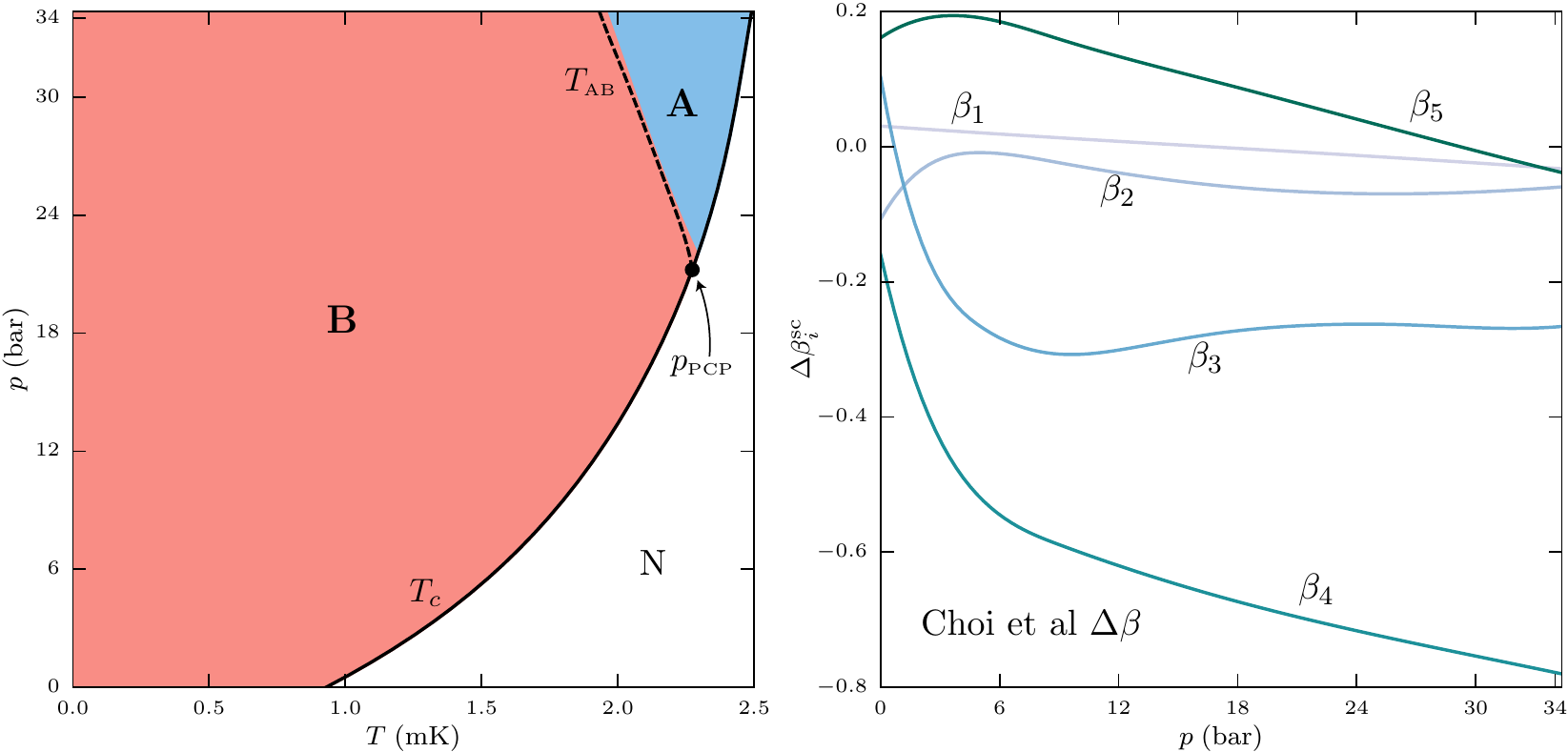}
\caption{(Left) Bulk phase diagram with lines showing the measured phase transitions and shading showing 
         the calculated regions of phase stability based on GL theory. The $\Delta\betasc{i}$ coefficients 
         are from Choi et al\cite{cho07} and are plotted in the right panel.}
\label{choi_bulk_diagram}
\end{center}
\end{figure}

\begin{figure}[t]
\begin{center}
\includegraphics[width=1\linewidth]{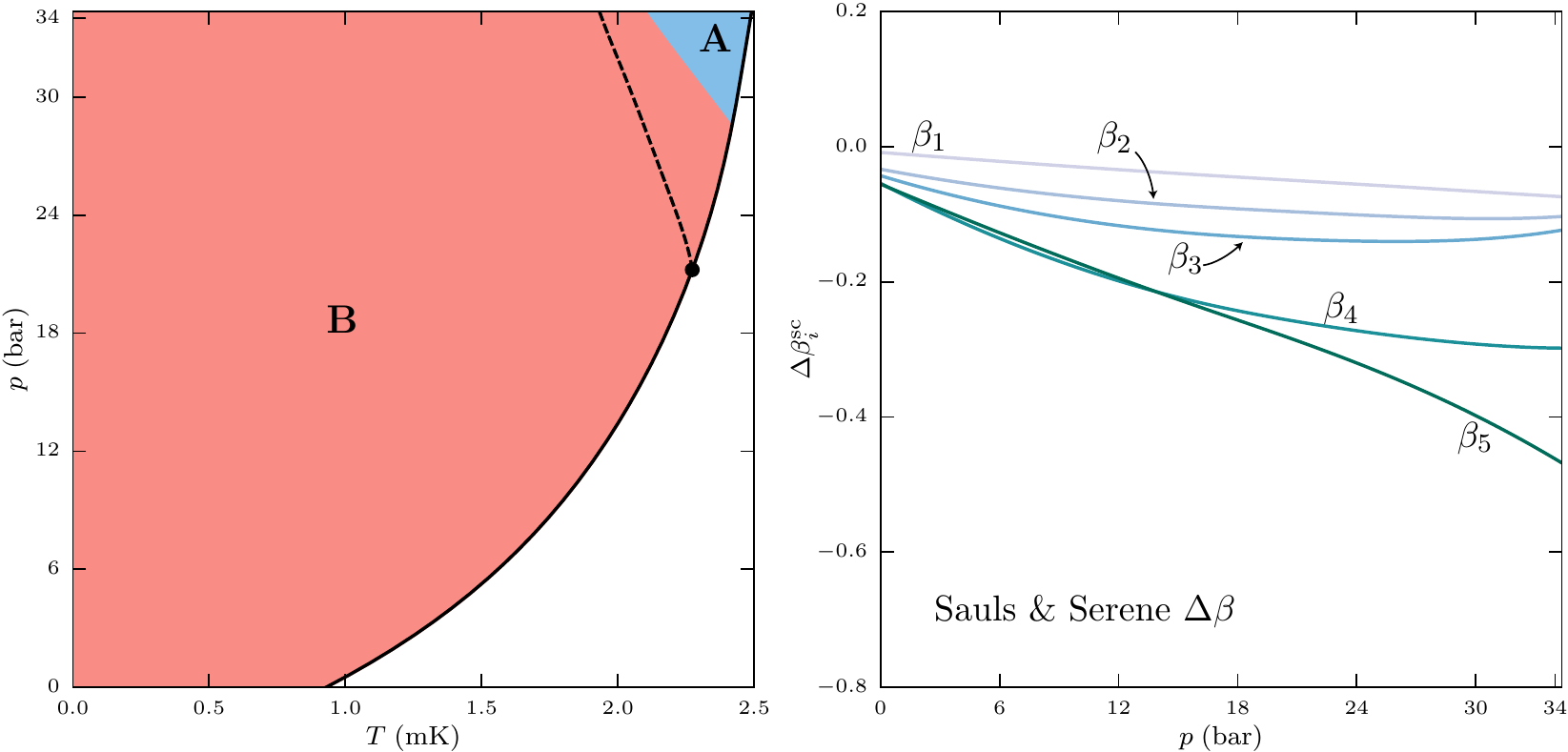}
\caption{(Left) Bulk phase diagram where the shaded regions represents the phases calculated from GL theory with 
	the $\Delta\betasc{i}$ of Sauls \& Serene \cite{sau81b}.
	These strong-coupling corrections are plotted in the right panel.}
\label{ss_bulk_diagram}
\end{center}
\end{figure}

The fourth order $\beta$ parameters that enter the GL free energy functional are modified 
by next-to-leading order corrections to the full Luttinger-Ward free energy functional \cite{rai76}.
These corrections scale as $\Delta\betasc{i}\sim\betawc{i}(T/T_F)$ near $T_c$.
Combining the $\Delta\betasc{i}$ with the weak-coupling coefficients in the bulk GL functional 
yields the critical pressure, $\pPCP{}$, above which the A phase is stable relative to the B phase. 
For $p>\pPCP{}$ the temperature scaling of the strong-coupling corrections relative to the weak-coupling $\beta$ 
parameters breaks the degeneracy in temperature between the A and B phases at the critical pressure and 
accounts for the pressure dependence of the A-B transition line, $T_{\text{AB}}(p)$, and thus an accurate 
bulk phase diagram \cite{wim15}. The resulting strong-coupling $\beta$ parameters are given by
\begin{align}
\label{eq-betas_sc-scaling}
\beta_i(T,p) 
&= \betawc{i}(p,T_c(p))+\frac{T}{T_c}\Delta\betasc{i}(p) 
\,.
\end{align}
Figure \ref{choi_bulk_diagram} shows the experimental bulk superfluid phase diagram as well as the 
phase diagram calculated from strong-coupling GL theory using $\Delta\betasc{i}$ coefficients obtained 
based on analysis of selected experiments by Choi et al. \cite{cho07}.
These $\beta$ coefficients differ substantively from those calculated from strong-coupling 
theory based on a quasiparticle scattering amplitude that accounts for the normal Fermi liquid 
properties of \He.
Figure \ref{ss_bulk_diagram} shows the bulk phase diagram calculated using the $\Delta\betasc{i}$ from 
Sauls \& Serene\cite{sau81b}. This set of $\beta$ coefficients has a higher polycritical pressure 
than experiment; however, the pressure dependence of the $\Delta\betasc{i}$ represents the expectation 
based on strong-coupling theory dominated scattering from ferromagnetic spin-fluctuations. 
Below $p = 12\mbar{}$ the $\Delta\betasc{i}$ are extrapolated to zero at a negative pressure 
corresponding to $T_c = 0$ \cite{wim15}.


\subsection{Boundary Conditions}
\label{subsec:boundary}
Confinement is represented in the GL theory through boundary conditions. For infinite, planar surfaces there are 
two limiting cases: maximal pairbreaking, due to the retroreflection of quasiparticles\cite{sau11}, 
and minimal pairbreaking, corresponding to specular reflection \cite{amb75}. For a surface on the $x-y$ plane 
with \He{} filling $z > 0$, maximal pairbreaking is defined within GL theory by
\begin{align}
A_{\alpha i}\big|_{z=0} &= 0\,\,\forall i\in\{x,y,z\}
\,,
\end{align}
while minimal pairbreaking is defined by
\begin{align}
A_{\alpha z}\big|_{z=0} &= 0\,, \nonumber \\
\nabla_z A_{\alpha x}\big|_{z=0} &= \nabla_z A_{\alpha y}\big|_{z=0} = 0
\,.
\end{align}

These boundary conditions may be extended by interpolating between the two extremes. In particular, 
Ambegaokar, de Gennes, and Rainer (AdGR) showed that diffuse scattering from an atomically rough 
surface leads to a GL boundary condition in which the transverse orbital components of the order 
parameter are finite at the surface, but extrapolate linearly to zero a distance $b_T = 0.54 \xi_0$ 
past the boundary. Thus, we introduce more general boundary conditions defined by
\begin{align}
A_{\alpha z}\big|_{z=0} &= 0\,, \nonumber \\
\nabla_z A_{\alpha x}\big|_{z=0} &= \frac{1}{b_T} A_{\alpha x}\big|_{z=0} \nonumber \\
\nabla_z A_{\alpha y}\big|_{z=0} &= \frac{1}{b_T} A_{\alpha y}\big|_{z=0} 
\label{eq:boundary}
\,,
\end{align}
where $b_T = b_T^\prime \xi_0$ is the extrapolation length. The parameter $b_T^\prime$ is allowed to 
vary from $b_T^\prime = 0$, maximal pairbreaking, to $b_T^\prime \to \infty$, minimal pairbreaking.
The film geometry consists of two infinite coplanar surfaces separated by a distance $D$ with \He{} 
filling the region between them. The boundary conditions in Eq. \ref{eq:boundary} are imposed at 
$z = \pm D/2$. 

\subsection{Extrapolating GL theory to low temperatures}\label{subsec:lengths}

\begin{figure}[t]
\begin{center}
\includegraphics[width=0.85\linewidth]{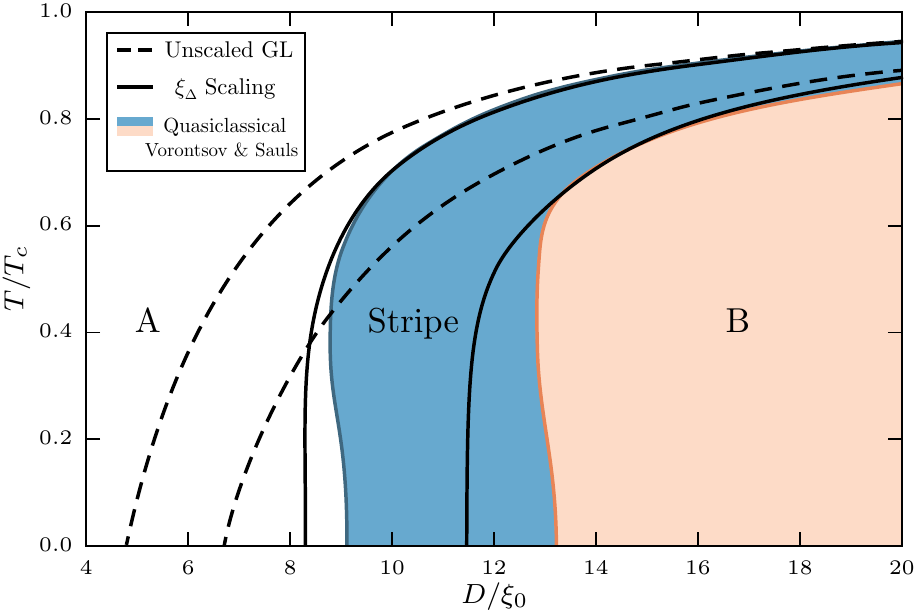}
\caption{Comparison of the phase diagrams calculated within weak-coupling quasiclassical theory
         (blue and orange lines), 
         weak-coupling GL theory (dashed lines), and 
         weak-coupling GL theory with $D$ rescaled by  
         $\xi_{GL}(T)/\xi_\Delta(T)$ (solid black lines).}
\label{scaling_diagram}
\end{center}
\end{figure}

Ginzburg-Landau theory is only expected to be accurate in the vicinity of $T_c$. 
This is easily seen in the order parameter amplitude, $\Delta^2\sim 1-T/T_c$, which 
varies linearly in $T$ down to $T=0$; whereas the weak-coupling BCS order parameter 
saturates at low temperatures. In confined \He, this difference is reflected in the 
characteristic length scale for variations of the order parameter, which in GL theory is
\begin{align}
\xi_{GL}(T)&=\left[\frac{7 \zeta(3)/20}{1-T/T_c}\right]^{1/2}
             \left(\frac{\hbar v_f}{2 \pi k_B T_c} \right) 
\,.
\end{align}
In weak-coupling BCS theory, the characteristic length scale is
\begin{align}
\xi_\Delta(T) &= \frac{\hbar v_f}{\sqrt{10} \Delta_B^{\text{\tiny BCS}}(T)} \,,
\end{align}
which is significantly larger than $\xi_{GL}(T)$ at low temperatures. In order to
more accurately extrapolate the spatial variations of the order parameter, as well
as the confinement phase diagram, to lower temperatures we rescale the 
film of thickness in the GL equations $D \to D(T)$ with
\begin{align}
D(T) &= D(T_c) \, \frac{\xi_{GL}(T)}{\xi_\Delta(T)}
\label{eq:lscale} 
\,,
\end{align}
where $D(T_c) = D$ is the thickness of the film and $D(T)$ is a rescaled thickness 
used within the GL theory calculation. 
Figure \ref{scaling_diagram} shows the effect of this rescaling on the weak-coupling 
GL theory phase diagram for the region of stability of the Stripe phase in comparison to 
the Stripe phase region obtained in weak-coupling quasiclassical theory \cite{vor07}.
Rescaling lengths in the GL theory in terms of $\xi_{\Delta}(T)$ gives a more accurate
representation of the confinement phase diagram than simple extrapolation of the GL 
results to low temperature. The deviations that remain reflect the non-locality of 
the quasiclassical theory for inhomogeneous phases for $T\ll T_c$. 

\section{Stripe phase}

The Stripe phase spontaneously breaks translational symmetry in the plane of the film. 
We assume it does so along the $x$ axis, leaving the order parameter translationally invariant 
along the $y$ direction. Broken translational symmetry leads to a new length scale, $L$, which 
is the half-period of the Stripe phase order parameter; $L$ is an emergent length scale, 
which varies with temperature, pressure, film thickness, and the surface boundary condition, 
and must be determined by numerical minimization of the GL free energy 
in parallel with the self-consistent determination of the order parameter.

\subsection{Order parameter}

The Stripe phase is predicted to be stable in superfluid \He{} films of thickness 
$D \sim 10 \xi_0$ \cite{vor07}. In weak-coupling theory this phase appears as a 
second order transition between the Planar and B phases, and for 
$D\lesssim D_{c_2}\approx 13\xi_0$, corresponds to a periodic array of degenerate 
B-phase domains separated by domain walls \cite{vor07}. 

\begin{figure}[t]
\begin{center}
\includegraphics[width=1\linewidth]{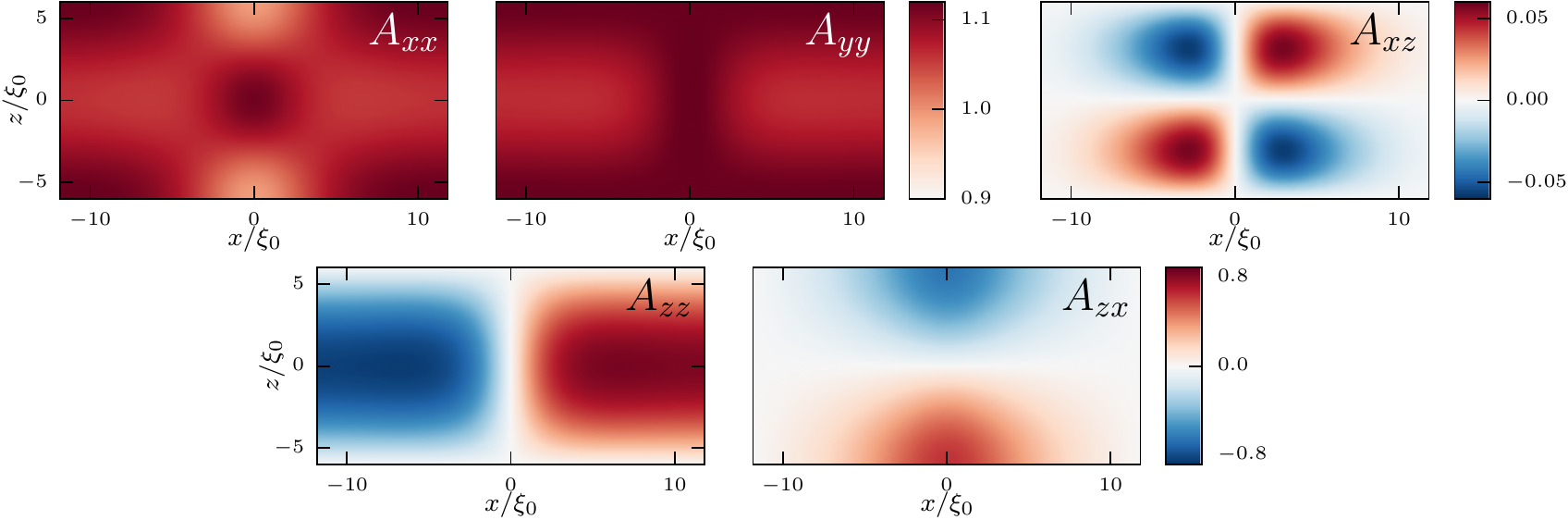}
\caption{Stripe phase order parameter for specular surfaces as functions $x$ 
         and $z$ for $D=12\xi_0$, $p=3\mbar$, $T=0.5T_c$, and calculated period $L\approx 23.6\xi_0$. 
         The amplitudes are scaled in units of the bulk B phase order parameter,
         $\Delta_B=\sqrt{|\alpha(T)|/6(\beta_{12}+1/3\beta_{345})}$.}
\label{order-parameter}
\end{center}
\end{figure}

For broken translational symmetry along the $x$ axis the residual symmetry of the Stripe phase 
is defined by the point group, 
\begin{eqnarray}
\mathsf{H} &=& \lbrace \mathrm{e},\, \mathsf{c}_{2x}^\mathsf{L} \mathsf{c}_{2x}^\mathsf{S} \rbrace \times 
\lbrace \mathrm{e}, \pi_{xz}^\mathsf{L}\pi_{xz}^\mathsf{S} \rbrace \ns\times\ns
\lbrace \mathrm{e}, \pi_{xy}^\mathsf{L}\pi_{xy}^\mathsf{S} \rbrace 
\nonumber
\\
&\times&
\lbrace \mathrm{e}, \pi_{xz}^\mathsf{L}\pi_{xz}^\mathsf{S} \rbrace \ns\times\ns
\lbrace \mathrm{e}, e^{i\pi} \mathrm{c}_{2z}^\mathsf{L} \rbrace \ns\times\ns \mathsf{T}\,,
\end{eqnarray}
where $\mathsf{c}_{2x}^\mathsf{L}$ is an orbital space $\pi$ rotation about the $x$ axis, 
$\pi_{xz}^\mathsf{S}$ is a spin space reflection about the $xz$ plane, and $\mathsf{T}$ is 
the operation of time reversal.
Based on this residual symmetry group we can simplify the form of the order parameter for the 
Stripe phase to
\begin{equation}
A(x,z) = 
\begin{pmatrix} 
A_{xx} & 0 & A_{xz} \\
0 & A_{yy} & 0 \\
A_{zx} & 0 & A_{zz} 
\end{pmatrix}
\,,
\end{equation}
where the remaining five components are functions of $x$ and $z$, and are all real due to time 
reversal symmetry.

The spatial dependences of the self-consistent order parameter components for the Stripe phase 
at pressure $p=3$ bar, $T/T_c=0.5$, thickness $D=12\xi_0$ with specular surfaces are shown in 
Fig. \ref{order-parameter}.
Note that the calculated half period is $L\approx 23.6\xi_0$, and that the 
dominant components are the diagonal elements, $A_{xx}$, $A_{yy}$ and $A_{zz}$. The latter
exhibits a domain wall separating degenerate B-like order parameters with $\mbox{sgn}(A_{zz})=\pm 1$.
The pair-breaking of $A_{zz}$ on the boundaries is alleviated by the large off-diagonal 
component, $A_{zx}$, at the junction with the domain wall. The remaining symmetry allowed amplitude,
$A_{xz}$, clearly exhibits the symmetry with respect to 
$\mathsf{c}_{2x}^\mathsf{L} \mathsf{c}_{2x}^\mathsf{S}$, but is smaller by an order of magnitude.

\begin{figure}[t]
\begin{center}
\includegraphics[width=1\linewidth]{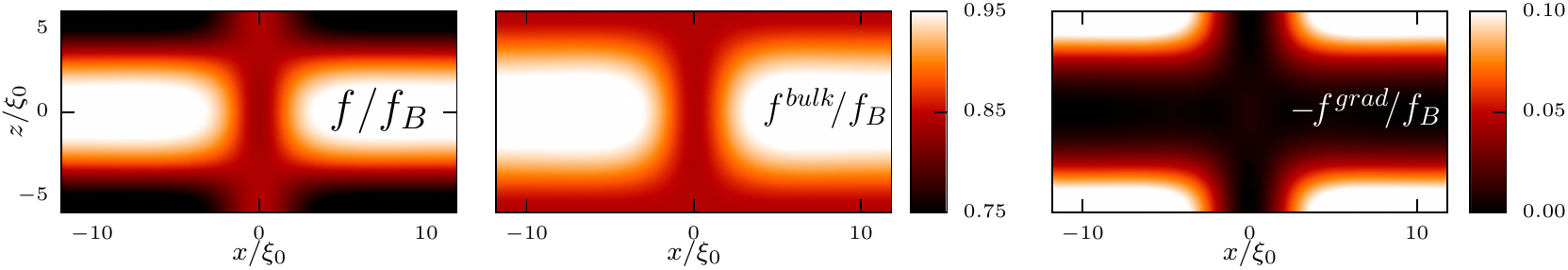}
\caption{Energy density of the Stripe phase with specular surfaces for $D=12\xi_0$, $p=3\mbar$, 
         $T=0.5T_c$, and calculated period $L\approx 23.6\xi_0$. The energy density $f$ is scaled 
         by the unconfined bulk energy density $f_B = \nicefrac{1}{2} \alpha(T)\,\Delta_B(T)^2 < 0$, 
         and is also shown separated into bulk and gradient contributions.}
\label{energy-density}
\end{center}
\end{figure}

The stability of the Stripe phase results from a tradeoff between the lowering of the energy
at junctions where the surfaces intersect the domain wall (note the gradient energy in 
Fig. \ref{energy-density}) and the cost in energy, away from the film surface, due to the 
suppression of the order parameter along the domain wall.
The total condensation energy density, with separate bulk and gradient energy densities, is
shown in Fig. \ref{energy-density}.

\subsection{Variational Model}

The magnitude of the half-period of the Stripe phase, $L$, is most easily determined using a variational 
form of the order parameter; $L$ is a minimum at the Stripe-Planar transition and diverges at the Stripe-B 
transition. At the Stripe-Planar transition, and for specular boundaries, $L$ may be derived from the
variational order parameter,
\begin{equation}
A(x,z) = 
\begin{pmatrix} 
\Delta_{xx} & 0 & 0 \\
0 & \Delta_{yy} & 0 \\
A_{zx} & 0 & A_{zz} 
\end{pmatrix}
\,,
\label{eq:varop}
\end{equation}
where $A_{zx}=-\Delta_{zx}\cos(\pi x/L)\sin(\pi z/D)$ and $A_{zz}=\Delta_{zz}\sin(\pi x/L)\cos(\pi z/D)$.
At the Stripe-Planar transition we assume that 
\begin{align}
\Delta_{yy} &= \Delta_{xx},\quad \Delta_{zx} \ll \Delta_{xx},\; \text{and }\; \Delta_{zz} \ll \Delta_{xx}.
\end{align} 
After spatially averaging and dropping terms greater than second order in $\Delta_{zx}$ and $\Delta_{zz}$
the resulting GL functional reduces to, 
\begin{align}
F_\mathrm{var} &= 2\alpha \Delta_{xx}^2 + 4\beta_P \Delta_{xx}^4  
- \frac{\pi^2 K_{23}\Delta_{zx}\Delta_{zz}}{2 D L} \\
& +\Delta_{zx}^2 \left\lbrace \frac{\alpha}{4} + \beta_P \Delta_{xx}^2
+\pi^2 \left(\frac{K_{123}D^2 + K_1 L^2}{4 D^2 L^2}\right) \right\rbrace \nonumber\\
& +\Delta_{zz}^2 \left\lbrace \frac{\alpha}{4} + \beta_{12} \Delta_{xx}^2
+\pi^2 \left(\frac{K_{1}D^2 + K_{123}L^2}{4 D^2 L^2}\right) \right\rbrace 
\nonumber
\,,
\end{align}
where $\beta_{ijk...} = \beta_i+\beta_j+\beta_k+... $, 
$K_{ijk...} = K_i+K_j+K_k+...$ and $\beta_P = \beta_{12}+ 1/2 \beta_{345}$ determines
bulk free energy of the Planar phase.
Minimizing $F_\mathrm{var}$ with respect to $\Delta_{xx}^2$ gives,
\begin{equation}
\Delta_{xx}^2 = \frac{|\alpha|}{2 \beta_P} - \frac{\Delta_{zx}^2}{8} 
              - \frac{\Delta_{zz}^2\beta_{12}}{8\beta_{P} } 
\,.
\end{equation}
The reduced free energy functional then simplifies to
\begin{align}
\label{eq:varlphaeq}
F_\mathrm{var} &= -\frac{\alpha^2}{4 \beta_P}
- \frac{\pi^2 K_{23}\Delta_{zx}\Delta_{zz}}{2 D L} \\
& +\Delta_{zx}^2 \left\lbrace \pi^2 \left(\frac{K_{123}D^2 
  + K_1 L^2}{4 D^2 L^2}\right) \right\rbrace \nonumber\\
& +\Delta_{zz}^2 \left\lbrace \alpha \left(\frac{\beta_P-\beta_{12}}{4 \beta_P}\right)
+\pi^2 \left(\frac{K_{1}D^2 + K_{123}L^2}{4 D^2 L^2}\right) \right\rbrace 
\,.
\nonumber
\end{align}
The last three terms in Eq. \ref{eq:varlphaeq} determine when nonzero values of 
$\Delta_{zx}$ and $\Delta_{zz}$ are favorable and the Stripe-Planar instability occurs. 
At the instability
\begin{align}
\label{eq:varalpha}
\alpha(T) &= -\frac{\pi^2 \beta_P}{D^2 L^2 (\beta_P - \beta_{12})}
\left\lbrace -2 D L K_{23} \left(\frac{\Delta_{zx}}{\Delta_{zz}}\right) \right.\\
&+\ns \left.\left(D^2 K_{123} + L^2 K_1 \right) \left(\frac{\Delta_{zx}}{\Delta_{zz}}\right)^2
\ns+\ns (D^2 K_1 + L^2 K_{123})
\right\rbrace \,.
\nonumber
\end{align}
Minimizing $F_\mathrm{var}$ with respect to the ratio $\Delta_{zx}/\Delta_{zz}$ gives 
\begin{equation}\label{eq:zxzzrat}
\frac{\Delta_{zx}}{\Delta_{zz}} = \frac{D L K_{23} } {D^2 K_{123} + L^2 K_1} 
\,.
\end{equation}
Combining Eq. \ref{eq:zxzzrat} with Eq. \ref{eq:varalpha} yields the Planar-Stripe 
instability temperature, $T_{\text{PS}}$, as a function of $D$ and $L$. Optimizing $T_{\text{PS}}$ with
respect to the Stripe phase period yields,  
\begin{equation}
\label{eq:l2full}
L = \sqrt{\frac{K_{123}}{|K_{23}-K_1|}}\,D
\,,
\end{equation}
which for weak-coupling values of $K_1$, $K_2$, and $K_3$, reduces to $L = \sqrt{3} D$. 

\begin{figure}[t]
\begin{center}
\includegraphics[width=0.95\linewidth]{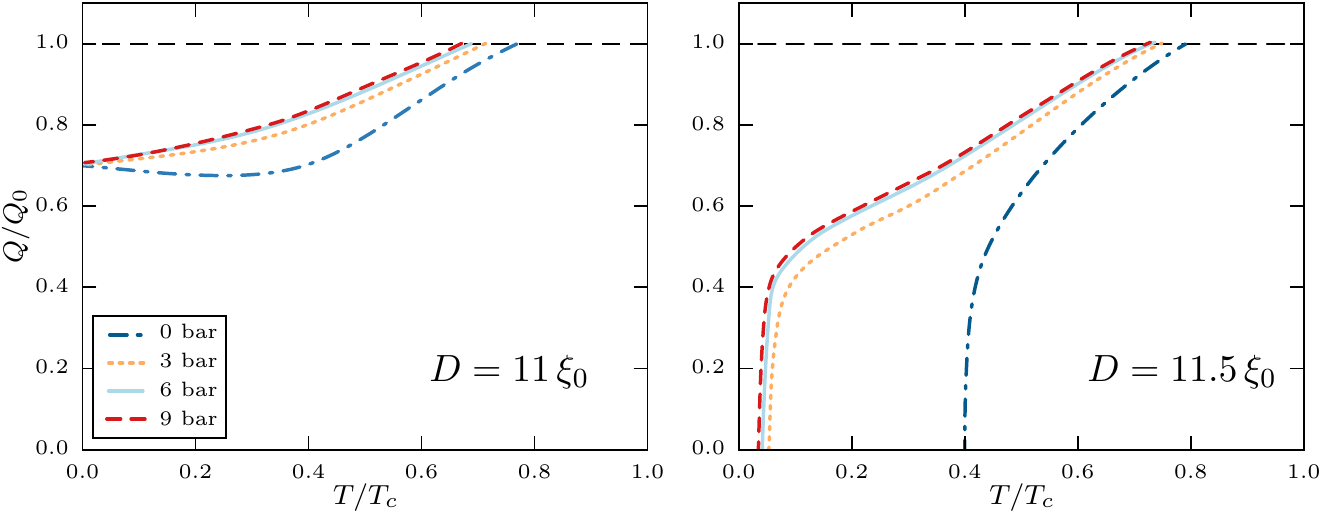}
\caption{Temperature and pressure dependence of the wavenumber $Q$ for film thicknesses
$D=11\,\xi_0$, with no Stripe to B transition (left panel), and 
$D=11.5\,\xi_0$, with a Stripe to B transition (right panel). The onset of the 
Stripe transition is based on the Planar-Stripe instability, i.e. omitting the A phase. 
}
\label{qplot}
\end{center}
\end{figure}

Although the Planar to Stripe transition is interrupted by a first-order transition to the 
A phase, the Stripe-Planar instability determines the scale of the half period, $L$, and the 
temperature region where the Stripe phase is expected to be stable.
The half-period defines the wavenumber, $Q_0 = \pi / \sqrt{3} D$, of the single-mode instability 
at $T_{\text{PS}}$. The wavenumber varies with the film thickness, $D$, and temperature. 
Figure \ref{qplot} shows the temperature dependence of $Q$ for two values of the film thickness 
starting from the Planar to Stripe instability at $T_{\text{PS}}$, i.e. omitting the A phase. 
The stability of the A-phase relative to the Planar phase changes the Stripe instability to a 
first-order transition at a lower temperature $T_{\text{AS}}$.
For $D=11\,\xi_0$ the stable region of Stripe phase persists to $T=0$, while for $D=11.5\,\xi_0$
there is a Stripe to B phase transition at a temperature, $T_{\text{SB}} < T_{\text{PS}}$. In 
both cases the wavenumber decreases ($L$ increases) as $T$ drops below $T_{\text{PS}}$, with
$Q\rightarrow 0$ ($L\rightarrow \infty$) as $T\rightarrow T_{\text{SB}}$. 
Strong coupling corrections to the free energy lead to a modest increase 
the period of the Stripe phase away from the Stripe to B transition; however, the transition 
temperature, $T_{\text{SB}}$, is sensitive to pressure (strong-coupling) as shown 
in the right panel of Fig. \ref{qplot}.

\section{Stripe Phase Stability}

\begin{figure}[t]
\begin{center}
\includegraphics[width=0.95\linewidth]{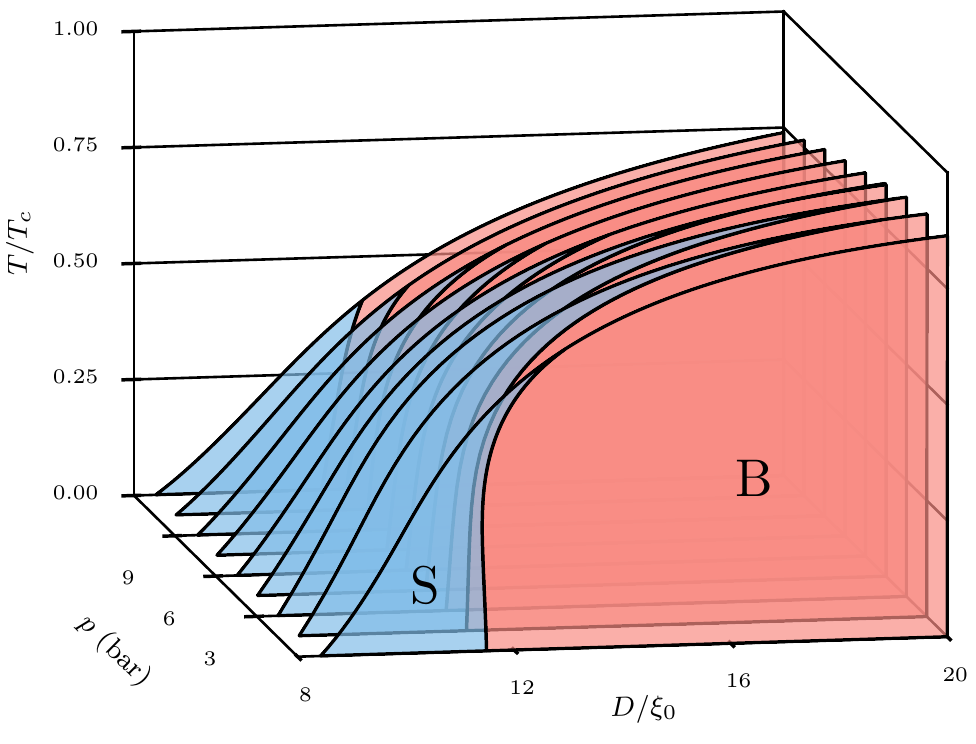}
\caption{Pressure-temperature-confinement phase diagram for the film with minimal pairbreaking 
         boundaries and experimental strong-coupling corrections. 
	 The A phase is stable everywhere not excluded by the Stripe and B phases.}
\label{pd-sc-3d}
\end{center}
\end{figure}

The most prominent effect of strong-coupling corrections to the weak-coupling BCS theory in bulk 
superfluid \He{} is the stability of the A phase above $\pPCP{}=21.22 \mbar{}$. 
In sufficiently thin films, the A phase is energetically stable relative to the B phase even in 
weak-coupling theory, and is degenerate with the Planar phase \cite{vor03,vor07}
Strong-coupling corrections favor the A phase over the Planar phase, leading to a stable A-phase 
in thin films at all pressures. Since the Stripe phase can be understood as a periodic array of
degenerate B phase domains separated by time-reversal invariant domain walls, one expects 
strong-coupling to favor the A phase near the Planar-Stripe instability line.
Indeed the A phase suppresses the Planar to Stripe instability temperature. However, 
the Stripe phase is found to be stable over a wide range of temperatures and pressures. 
 
Figure \ref{pd-sc-3d} shows the phase diagram for minimal pairbreaking (specular) surfaces at 
pressures from $0$ to $12\mbar{}$, with the Stripe phase onsetting at temperatures above 
$0.5 T_c$. The accuracy of the strong-coupling GL theory is expected
to diminish at very low temperatures; therefore we show results for low and intermediate pressures 
for which the A- to Stripe transition onsets above $0.5 T_c$.
Note that at $T=0$ the strong-coupling GL corrections vanish, and the phase boundaries are determined 
by weak-coupling theory at $T=0$ and thus pressure independent. This is an artefact of the 
temperature scaling of the strong-coupling GL parameters. It is known that there are  
residual strong-coupling corrections at the few percent level in the limit $T=0$ \cite{ser83}.

A striking difference between the two sets of strong-coupling $\beta$ parameters shown in Figs. 
\ref{choi_bulk_diagram} and \ref{ss_bulk_diagram} is evident at low pressures. The $\Delta\betasc{i}$
from Choi et al. \cite{cho04} are non-monotonic between $p=0$ and $p=12$ bar, which leads to maximal 
stability of the Stripe phase at $p \approx 3\mbar{}$. In contrast the theoretically calculated
strong-coupling corrections are monotonic functions of pressure and predict maximal stability 
of the Stripe phase at $p=0 \mbar{}$ and decreasing stability with increasing pressure.

\subsection{Pressure-Temperature Phase Diagram}

\begin{figure}[t]
\begin{center}
\includegraphics[width=1.0\linewidth]{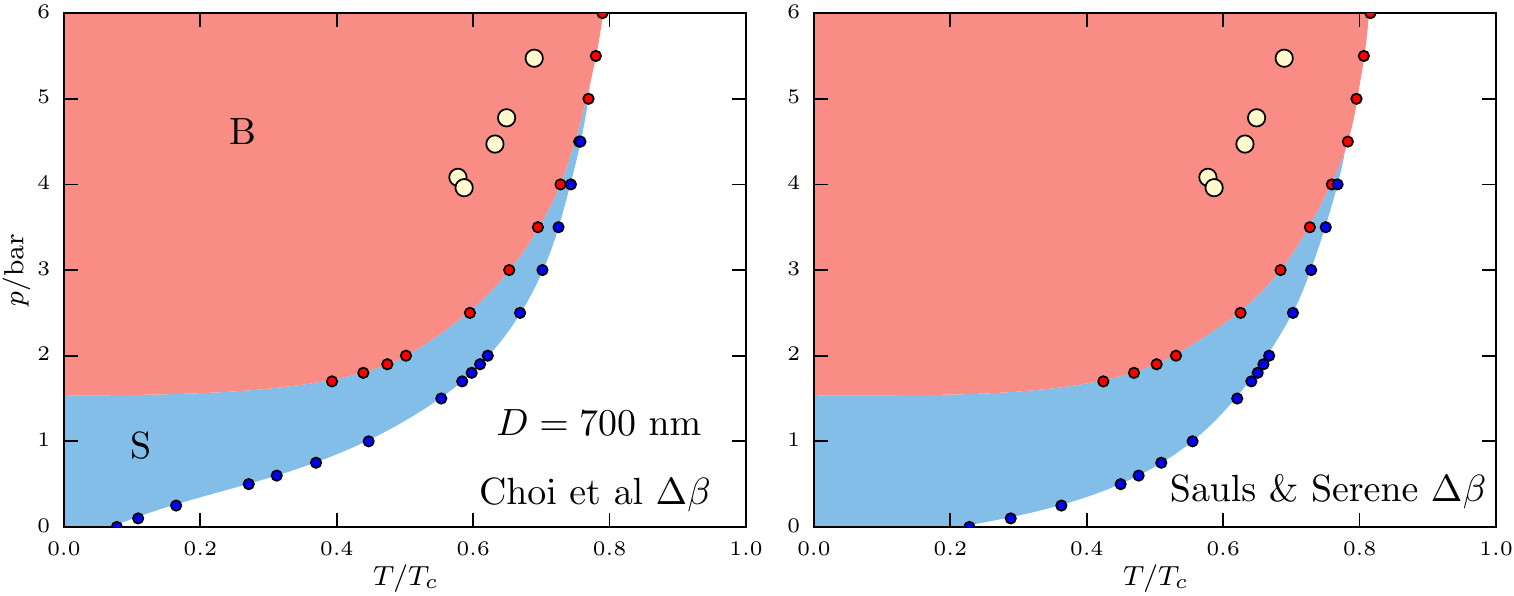}
\caption{Pressure-temperature phase diagram for a film of thickness $D = 700 \nm{}$ with minimal 
         pairbreaking (specular) boundary conditions.
	 The A phase is stable everywhere in the white region below the bulk transition temperature.
	 The larger yellow circles are data for the A-B transition based on NMR from Levitin et al 
         obtained with $^4$He preplating \cite{lev13}.}
\label{700nm-pd}
\end{center}
\end{figure}

Although a number of experiments have been reported on superfluid \He{} in planar geometries, of particular 
interest are those involving slabs of thickness $D \approx 700 \nm{}$ and $D \approx 1080 \nm{}$, which 
are in the range of confinement where the Stripe phase is expected to be stable.
Levitin et al. \cite{lev13} (RHUL group) used transverse NMR frequency shifts to determine transition 
temperatures in these cells. They did not find NMR evidence of the Stripe phase. These experiments were 
done both with and without preplating the surfaces of the slab with $^4$He, the presence of which greatly 
increases the specularity of the surface. Without the $^4$He present, the RHUL group reported large 
suppression of the onset of the superfluid transition - a suppression larger than that predicted 
theoretically for maximally pairbreaking retro-reflective surface scattering. The explanation or origin of 
this anomalous suppression is currently lacking. Thus, we focus on the measurements done with $^4$He 
preplating, which exhibit minimal $T_c$ suppression, and may be modeled theoretically with minimal 
pairbreaking boundary conditions (specular scattering).

Calculations of the phase diagram for $D = 700 \nm{}$ are shown in Fig. \ref{700nm-pd}. The A phase 
onsets at the bulk $T_c$. There is an A to Stripe transition followed by the Stripe to B transition.
For both sets of  strong-coupling $\beta$ parameters, the Stripe phase is predicted to be stable 
at low pressures and at experimentally accessible temperatures. Although the stability of the A phase is 
maximal with specular boundary conditions, the calculated A-B or A-S phase transition occurs at significantly 
higher temperature than that reported by the RHUL group. The discrepancy is sufficiently large that 
it is well outside uncertainties in the magnitude of the strong-coupling parameters based bulk A- and 
B phase free energies. Based on our calculations accessing the Stripe phase
would be optimal for pressures between $p=1$ and $p=1.5$ bar.

\begin{figure}[t]
\begin{center}
\includegraphics[width=1.0\linewidth]{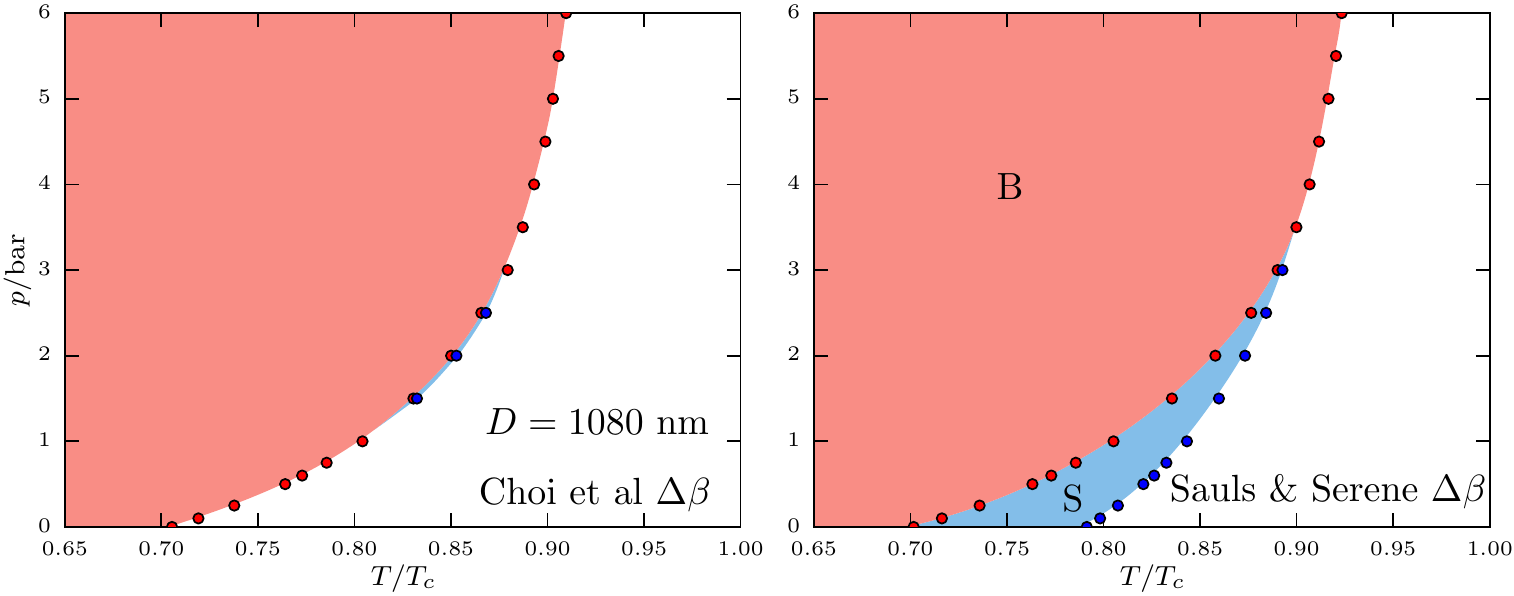}
\caption{Pressure-temperature phase diagram for a film of thickness $D = 1080 \nm{}$ with minimal 
         pairbreaking boundaries. The A phase is stable everywhere not excluded by the Stripe and 
         B phases.}
\label{1080nm-pd}
\end{center}
\end{figure}

For the thicker slab geometry, $D = 1080 \nm{}$, shown in Figure \ref{1080nm-pd}, the Stripe phase is 
predicted to have a negligible region of stability in the pressure-temperature plane based on the 
$\beta$ parameters from  Choi et al. \cite{cho04}, and only a small window of stability at the lowest 
pressures based on the theoretically calculated strong-coupling parameters. 

\subsection{Effects of Surface Conditions on the Phase Diagram}

\begin{figure}[t]
\begin{center}
\includegraphics[width=0.8\linewidth]{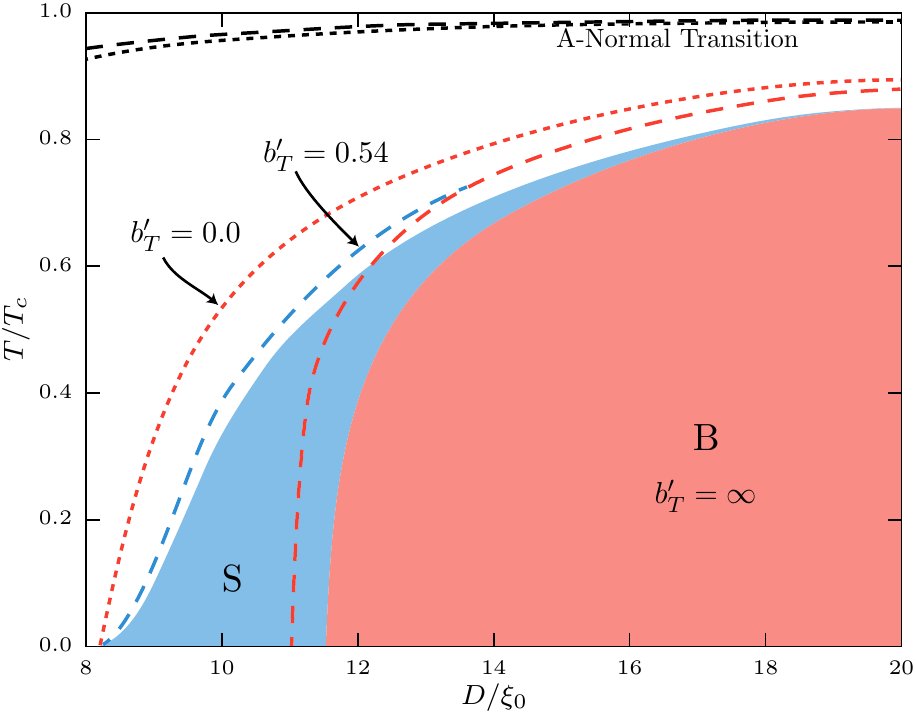}
\caption{Temperature-confinement phase diagram for films at $p = 3 \mbar{}$ with the Choi et al.
         strong-coupling corrections. Results for three boundary conditions are shown: minimal 
         pairbreaking, $b_T^\prime \to \infty$ (solid); diffuse, $b_T^\prime = 0.54$ (dashed); and 
         maximal pairbreaking, $b_T^\prime = 0$ (dotted). For diffuse and maximal pairbreaking, the 
         suppression of the A to Normal phase transitions are also shown.}
\label{boundary-comp}
\end{center}
\end{figure}

We use the variable boundary conditions in Eqs. \ref{eq:boundary} to investigate the sensitivity 
of the Stripe phase to surface disorder. 
Figure \ref{boundary-comp} shows the temperature-confinement phase diagram at $p = 3\mbar{}$ for
maximal ($b_T^\prime=0$), diffuse ($b_T^\prime=0.54$), and minimal ($b_T^\prime=\infty$) pairbreaking 
boundary conditions. 
Maximal stability of the Stripe phase occurs for minimal pairbreaking, i.e. specular surfaces, as
shown by the blue region of stable Stripe phase.
Note that for diffuse scattering the region of Stripe phase stability does not differ significantly 
from that for specular boundary scattering.
Conversely, for maximal pairbreaking the Stripe phase exists only in the vicinity of $T=0$.

\section{NMR Signatures of the Stripe Phase}

Nuclear magnetic resonance (NMR) spectroscopy of the \He\ order parameter
is based on resonance frequency shifts originating from the Cooper pair contribution 
to the nuclear magnetic dipole energy, $\Delta\Omega_D = \int_V d^3r\,f_D[A]$, which
evaluated to leading order in $A$ is
\begin{align}
f_{D} = g_D \left( |Tr{A}|^2 + Tr{AA^*} \right) 
\,,
\label{eq-dipole_energy}
\end{align}
where $g_{D}=\nicefrac{\chi}{2\gamma^2}\Omega_A^2/\Delta_A^2$ is the nuclear dipole coupling,
$\gamma$ is the \He\ nuclear gyromagnetic ratio, $\chi$ is the nuclear magnetic susceptibility 
of normal \He, and $\Omega_A$ is the A phase longitudinal NMR resonance frequency. The 
dipole energy, of order $g_{D}\Delta_A^2$, lifts the degeneracy of 
relative rotations of the spin- and orbital state of the Cooper pairs. 

NMR spectroscopy is based on the NMR frequency shift, $\Delta\omega = \omega-\omega_L$, 
resulting from the dipolar torque acting on the total nuclear magnetization. The shift
depends in general on the orientation of the NMR field, $\vH$, the initial tipping angle,
$\beta$, generated by the r.f. pulse, and particularly the spin- and orbital structure of
the order parameter. We use the reduction of Leggett's theory of NMR in \He\ 
proposed by Fomin \cite{fom78}, valid for intermediate magnetic fields, 
$\Omega_A \ll \omega_L \ll \Delta$, where $\omega_L=\gamma H$ is the Larmor frequency \cite{fom78}. 
The key approximation is the first inequality which provides a separation of ``fast'' and ``slow'' 
timescales for the spin dynamics. The second inequality allows us to neglect the deformation
of the order parameter by the Zeeman field.
Similarly, for inhomogeneous states we use the separation of length scales for spatial variations 
of the Stripe phase, of order $L\sim D\approx 1\,\mu\mbox{m}$, both small compared to the dipole 
coherence length, $\xi_D\equiv \sqrt{g_D/K_1}\approx 20\,\mu\mbox{m}$.
The spin degrees of freedom of the order parameter cannot vary on length scales shorter 
than the dipole coherence length $\xi_D$. Thus, for $L\ll\xi_D$ the nuclear spin dynamics 
is determined by the spatially averaged dipole energy. An exception to this spatial averaging
occurs near the Stripe-B transition where the period of the Stripe phase diverges. In this
limit the dipolar energy varies on sufficiently long spatial scales that the spin dynamics is
determined by a spatially varying dipolar potential. 
Combined with Fomin's formulation, the separation in scales for spatial variations of the 
orbital and spin components of the order parameter allows us to  
calculate the nonlinear NMR frequency shifts for the inhomogeneous phases of the thin film 
as described in Ref. \cite{wim15}.

\subsection{Translationally invariant Planar-distorted B phase}

For non-equal-spin pairing (non-ESP) states, e.g. the polar distorted B phase or the Stripe phase, the 
nuclear magnetic susceptibility, $\chi$, is suppressed relative to that of normal \He, $\chi_N$. 
For all non-ESP phases, including the Stripe phase, the susceptibility can be expressed as
\begin{equation}
\chi_B = \frac{\chi_N}{1+2\,g_z /\chi_N {(\langle \Delta_{zx}^2 \rangle 
       + \langle \Delta_{zz}^2 \rangle)}} 
\,.
\label{eq:chi_b}
\end{equation}

For a non-ESP superfluid phase of a \He\ film with the magnetic field $\vec{H}||\vz$, 
for both the B and Stripe phases, there are two possible dipole orientations corresponding to 
different local minima in the dipole energy \cite{bun93}. The first orientation is 
a minimum of the dipole energy and has positive frequency shift, which following Levitin et al. 
we denote as the $\mathrm{B}^+$ state in the case of the translationally invariant 
B phase. The frequency shift for the $\mathrm{B}^+$ state is obtained as
\begin{widetext}
\begin{align}
&\omega\Delta\omega^{+} = \frac{\gamma^{2}}{\chi_{B}}g_{D} 
\times 
\begin{cases}
\frac{\Dxx^2-\DxxDzz^2}{\Dxx} + 2\left( \frac{\DxxDzz^2}{\Dxx} - \Dzz \right)\cos\beta\,,
& \cos\beta \geq \cos\beta^*\,, \\
-\Dxx-\DxxDzz - 2\left\langle\vphantom{\frac{1}{2}} (A_{xx}+A_{zz})^2 \right\rangle\cos\beta\,,
& \cos\beta < \cos\beta^* \,,
\end{cases}
\end{align}
\end{widetext}
where $\left\langle ... \right\rangle=(1/V)\int_V\,d^3R\ldots$ denotes spatial averaging, and
\begin{equation}
\cos\beta^* = \frac{1}{2}\left(\frac{\DxxDzz-2\Dxx}{\DxxDzz+\Dxx}\right)
\end{equation} 
is the critical angle.  

Axial symmetry of the Planar-distored B phase implies $\Dyy = \Dxx$; thus, only $\Dxx$, $\Dzz$, 
and $\DxxDzz$ are non-zero. This NMR resonance is analogous to the Brinkman-Smith mode in 
bulk \He{}-B, but with a positive frequency shift at small tipping angle and a shifted critical 
angle.

The translationally invariant, but meta-stable, $\mathrm{B}_-$ state, corresponds to a 
minimum of the dipole energy, and has a frequency shift given by
\begin{align}
\omega\Delta\omega^{-} &= \frac{\gamma^{2}}{\chi_{B}}g_{D} 
\left\lbrace\vphantom{\wpreB{}} -\left(\Dxx + 2\Dzz\right)\cos\beta\right\rbrace 
\,.
\end{align}
This mode has a negative frequency shift at small tipping angles and, unlike the $\mathrm{B}_+$ state,
has no critical angle, and therefore no deviation from cosine tipping angle dependence.
The tipping angle dependences of both Planar-distorted B phase states are shown in 
Fig. \ref{nmr} plotted as a function of $\cos\beta$.
The positive (negative) shift at small tipping angle is the signature of the of the $\mathrm{B}^{+}$ 
($\mathrm{B}^{-}$) state in the NMR spectra of the RHUL group \cite{lev13}. These identifications
are confirmed by nonlinear NMR measurements \cite{lev13b} showing both the pure cosine tipping angle 
dependence of $\omega\Delta\omega$ for the $\mathrm{B}^{-}$ state, and the ``kink'' in the shift at 
the critical angle $\beta^*$ for the $\mathrm{B}^{+}$ state. Note that for $D=12\xi_0$ at $p=3$ bar
there is a small slope to the positive shift for $\cos\beta>\cos\beta*$.

\subsection{Nonlinear NMR shifts for the $\mathrm{S}^{\pm}$ Stripe phases}

\begin{figure}[t]
\begin{center}
\includegraphics[width=1\linewidth]{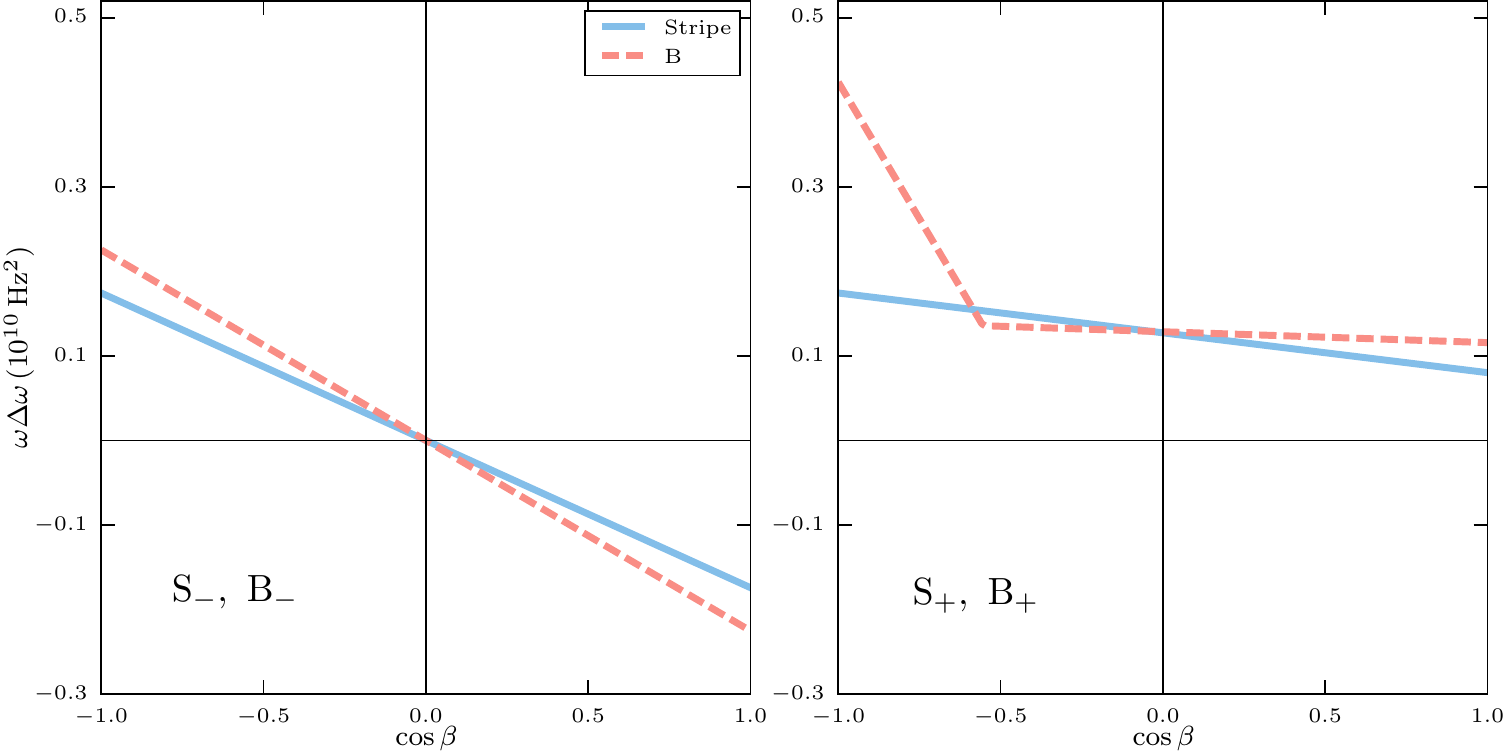}
\caption{Transverse NMR frequency shifts as a function of tipping angle $\beta$ at 
	$D=12\xi_0$, $p=3\mbar{}$, and $T=0.5\,T_c$, with minimal pairbreaking 
        for the $\mathrm{B}^{\pm}$ and $\mathrm{S}^{\pm}$ states.}
\label{nmr}
\end{center}
\end{figure}

The breaking of both translational and rotational symmetry in the plane of the film by the
Stripe phase leads to a qualitatively different transverse NMR frequency shift for the 
Stripe phase with relative spin-orbit rotation corresponding to a 
minimum of the dipole energy, i.e. the $\mathrm{S}^+$ state,
\begin{widetext}
\begin{align}
\omega\Delta\omega^{+} =
\frac{1}{2}\frac{\gamma^{2}}{\chi_{B}}g_{D} 
\left\lbrace\vphantom{\wpreB{}} \DxxPDyy \right. 
-\left. \vphantom{\wpreB{}}\left\lbrack \DxxMDyy + 8\Dzz 
-4
  \left(\Dxz + \Dzx\right)\right\rbrack\cos\beta  \right\rbrace 
\,.
\end{align}
\end{widetext}
The $\mathrm{S}^{+}$ phase is distinguished with respect to both the bulk B phase and the 
Planar-distorted $\mathrm{B}^+$ phase by the absence of a critical tipping angle. This results 
from spatial averaging over the period of the Stripe phase which contains equal volumes of 
$A_{zz} > 0$ and $A_{zz}<0$ giving $\langle A_{xx} A_{zz} \rangle = \langle A_{yy} A_{zz} \rangle = 0$.

By contrast the frequency shift of the metastable $\mathrm{S}^{-}$ phase does not differ 
substantially from that of the $\mathrm{B}^{-}$ phase,
\begin{widetext}
\begin{align}
\omega\Delta\omega^{-} 
= \frac{1}{2}\frac{\gamma^{2}}{\chi_{B}}g_{D} 
\left\lbrace\vphantom{\wpreB{}} \DxxMDyy(1+\cos\beta) \right. 
-\left. \vphantom{\wpreB{}}\left\lbrack 2\Dxx + 2\Dyy +8\Dzz - 4\left(\Dxz 
+ \Dzx\right)\right\rbrack\cos\beta\right\rbrace 
\,.
\end{align}
\end{widetext}
Note that the constant term in the shift for the $\mathrm{S}^{-}$ state proportional to the 
average $\DxxMDyy$ is absent for the $\mathrm{B}^-$ state; however, this constant shift is 
negligibly small. Figure \ref{nmr} shows the comparison between the translationally invariant
$\mathrm{B}^{\pm}$ NMR shifts and those for the corresponding stable and metastable 
$\mathrm{S}^{\pm}$ Stripe phases. 
The primary NMR signature of the Stripe phase is the positive shift with an offset, a finite slope
and the absence of critical angle. This signature clearly differentiates the $\mathrm{S}^{+}$
phase from the $\mathrm{B}^{\pm}$ states and the A phase.

\section{Summary and Outlook}
By formulating a GL theory that incorporates pressure and temperature dependent strong-coupling 
corrections, combined with temperature dependent rescaling of the confinement length, $D$, 
we have greatly expanded the region of applicability of GL theory for calculations of the 
properties of confined superfluid \He{}.
Strong-coupling corrections expand the region of stability of
the A phase and decrease the region of stability of the Stripe phase; however, the Stripe phase 
remains stable in a large region of pressure, temperature, and confinement. 
The stability of the Stripe phase is insensitive to diffuse surface scattering;  
the phase diagram for specular and fully diffusive scattering 
predict the Stripe phase to occur in nearly equivalent regions of the phase diagram.
Nonlinear NMR measurements are probably the best means of detecting the Stripe phase. The 
NMR signatures - positive shift with no critical angle  - differentiates the $\mathrm{S}^{+}$ 
phase from the $\mathrm{B}^{\pm}$ and A phases. 

\begin{acknowledgements}
The research of JJW and JAS was supported by the National Science Foundation (Grants DMR-1106315 and DMR-1508730).
\end{acknowledgements}

\end{document}